\documentclass{article}
\usepackage{spconf,amsmath,graphicx}
\usepackage{algorithm,algpseudocode}
\usepackage{booktabs,multirow}

\usepackage{setspace}

\title{Tiny Transducer: A highly-efficient Speech recognition Model on Edge Devices  }
\name{Yuekai Zhang$^{1,2}$\sthanks{Work performed during internship at Tencent.}, Sining Sun$^1$\sthanks{The first two authors contributed equally to this work},  Long Ma$^1$}
\address{Tencent, JHU}
\address{$^1$Tencent Technology Co.,Ltd, Beijing, China\\
	$^2$The Johns Hopkins Univeristy, Baltimore, MD, USA\\
	yzhan400@jhu.com,   \{siningsun,malonema\}@tencent.com}

\begin{document}
%\ninept
%
\linespread{0.4}
\maketitle

\begin{abstract}

 This paper proposes an extremely lightweight phone-based transducer model with a tiny decoding graph on edge devices. First, a phone synchronous decoding (PSD) algorithm based on blank label skipping is first used to speed up the transducer decoding process. Then, to decrease the deletion errors introduced by the high blank score, a blank label deweighting approach is proposed. To reduce parameters and computation, deep feedforward sequential memory network (DFSMN) layers are used in the transducer encoder, and a CNN-based stateless predictor is adopted. SVD technology compresses the model further. WFST-based decoding graph takes the context-independent (CI) phone posteriors as input and allows us to flexibly bias user-specific information.   Finally, with only 0.9M parameters after SVD, our system could give a relative 9.1\% - 20.5\% improvement compared with a bigger conventional hybrid system on edge devices.

\end{abstract}
%

%\vspace{-1mm}

\begin{keywords}
Transducer, on-device model,  phone synchronous decoding
\end{keywords}
%\vspace{-4mm}
\section{Introduction}
\label{sec:intro}

%\vspace{-2mm}
Recently, end-to-end (E2E) models \cite{kim2017joint,wang2019exploring,guo2020recent,gulati2020conformer,li2020developing} for automatic speech recognition (ASR) have become popular in the ASR community. Comparing with conventional ASR systems \cite{ghoshal2013sequencediscriminative,peddinti2015time}, including three components: acoustic model (AM), pronunciation model (PM), and language model (LM), E2E models only have a single end-to-end trained neural model but with comparable performance with the conventional systems. Thus, E2E models are gradually replacing the traditional hybrid models in the industry \cite{li2020developing,sainath2020streaming}. \\
Another research line focuses on deploying ASR systems on devices such as cellphones, tablets, and embedded devices \cite{sainath2020streaming,he2019streaming,park2018fully,wang2020cascade}. %Instead of shrinking the conventional hybrid systems components, directly deploying the E2E models on devices is another attractive option. 
However, deployment of E2E models on devices remains several challenges: first, on-device ASR tasks usually require a streamable E2E model with low latency. Popular E2E models such as attention-based encoder-decoder (AED) \cite{dong2018speech,karita2019comparative,luo2020simplified} have shown state-of-the-art performance on many tasks, but the attention mechanism is naturally unfriendly to online ASR. Second, the customizable ability is desired in many on-device ASR scenarios. The model should have a promising performance on user-specific information such as contacts' phone numbers and favorite song names.
In \cite{zhao2019shallow}, shallow fusion is combined with E2E models' prediction during decoding.  In \cite{sim2019personalization}, text-to-speech (TTS) technology is utilized to generate training samples from text-only data. However, they all need to retrain the acoustic model or language model (LM).
Finally, especially on edge devices where the memory and computing resources are highly constrained, ASR systems have to be very compact. (e.g., Embedded devices for vehicles could only attribute low memory and computing budget to ASR.) \\ 
To satisfy the above requirements, we present a highly-efficient ASR system, suitable for ASR tasks with insufficient computing resources. Our proposed system consists of a lightweight phone-based speech transducer and a tiny decoding graph.  The transducer converts speech features to phone sequences. The decoding graph, composing of a lexicon and a grammar FST , named LG graph, maps phone posteriors to word sequences. On the one hand, compared with conventional senone-based acoustic modeling, phone-based speech transducer simplifies the acoustic modeling process. On the other hand, combining with the LG graph will easily fuse language model or bias user-special information into the decoding graph. \\
Within our proposed architecture, we first adopt a phone synchronous decoding (PSD) algorithm based on transducer with blank skipping strategy, improving decoding speed dramatically  with no recognition performance drop. Then, to alleviate the deletion error caused by the over-scored blank prediction, we propose a blank label deweighting approach during speech transducer decoding, which can reduce the deletion error significantly in our experiments. 
To reduce model parameters and computation, a deep feedforward sequential memory block (DFSMN) is used to replace the RNN encoder, and a casual 1-D CNN-based (Conv1d) stateless predictor \cite{ghodsi2020rnn,weng2019minimum} is adopted. Finally, we apply the singular value decomposition (SVD) to our speech transducer to further compress  the model. Our tiny transducer could achieve a promising performance with only 0.9M parameters.\\

\section{Tiny Transducer}
\label{sec:format}
\begin{figure}[H]
  \centering
  \includegraphics[width=0.7\linewidth]{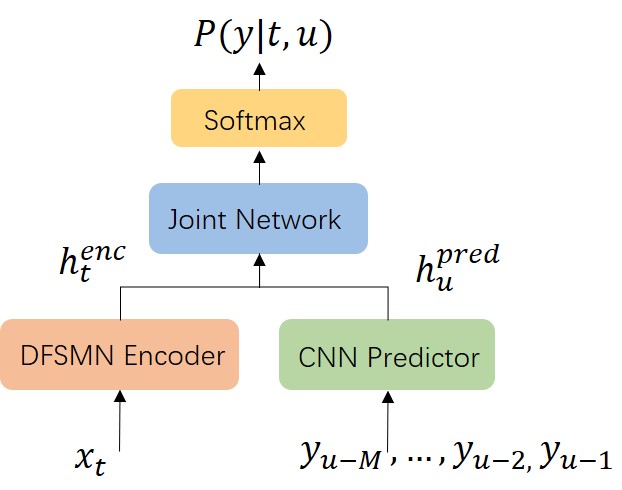}
  %\vspace{-5mm}
  \caption{The Architecture of Transducer Model}
  \label{fig:speech_production}
  \vspace{-3mm}
\end{figure}

%\vspace{-3mm}
RNN-T model is proposed in \cite{graves2012sequence} as an improvement of connectionist temporal classification (CTC) \cite{graves2006connectionist}, which removes the strong prediction independence assumption of CTC. RNN-T includes three parts: an encoder, a predictor, and a joint network. Traditionally, both encoder and predictor consist of a multi-layer recurrent neural network such as LSTM, resulting in high computation on devices. In this work, DFSMN-based encoder and a casual Conv1d stateless predictor are used to achieve efficient computation on devices. Fig~\ref{fig:speech_production} illustrates the architecture of our transducer model. 

\vspace{-5mm}

%%\vspace{-4mm}

\subsection{Streamable DFSMN Encoder}
\label{ssec:subhead}
%%\vspace{-3mm}
Due to the limited computation resources, the popular streaming architecture, LSTM, is replaced with a DFSMN layer. DFSMN combines FSMN \cite{zhang2015feedforward} with low-rank matrix factorization \cite{zhang2016compact,zhang2018deep} to reduce network parameters.% A skip connection is also used to address the gradient vanishing problem.
To keep a good trade-off between model accuracy and latency, we set the number of left context frames in the DFSMN layer as eight, and the right context includes two frames. In this way, the deeper layers would have a wider receptive field with more future information. Additionally, two CNN layers with stride size two each layer are inserted before the DFSMN layers to perform subsampling, which leads to  four times subsampling. 

%%\vspace{-4mm}

\subsection{Casual Conv1d Stateless Predictor}
%%\vspace{-2mm}
\label{ssec:conv1d predictor}
The predictor network only has one casual Conv1d layer. It takes $M$ previous predictions as input. We set M to four in our experiments. Formally, the predictor output at step $u$ is 
%%\vspace{-2mm}
\begin{equation}
    \mathbf{h}_u^{pred}={\rm Conv1d}({\rm Embed}(y_{u-M}, ..., y_{u-1}))
\end{equation}
where Embed() maps the predicted labels to the corresponding embeddings. 
Fig~\ref{fig:speech_production} also shows our Conv1d predictor.

%\vspace{-5mm}
\section{Decoding WITH TINY transducer}
\label{sec:print}
%\vspace{-3mm}
In this work, we choose to use CI phones as prediction units. Combining with a traditional WFST decoder allows us to flexibly inject biased contextual information into the decoder graph without retraining the acoustic model and LM model. During the decoding process, the CI phone probability posteriors from the transducer model would be the WFST decoder's input. Our WFST decoder includes two separate WFSTs: lexicons (L), and language model, or grammars (G). The final search graph ($LG$) can be presented as follow:
%\vspace{-2mm}
\begin{align}
    LG = min(det(L \circ G))
\end{align}
%%\vspace{-2mm}
where $min$ and $det$ represent determinize and minimize operations, respectively.
To further speed up the decoding process and reduce parameters, we introduce our PSD algorithm and SVD technology in this section. 

\subsection{Phone Synchronous Decoding with Blank Skipping}
\label{ssec:subhead}
%\vspace{-1mm}
PSD algorithm is first used in \cite{chen2016phone} to speed up the decoding and reduce the memory usage with CTC lattice. A CTC model's peaky posterior property allows the PSD algorithm to ignore blank prediction frames and compress the search space. We found the same peaky posterior property also exists in an RNN-T model. With transducer lattice, most frames are aligned with blank symbols. Motivated by this, we presented a PSD algorithm based on RNN-T lattice. We introduce our PSD method below.\\
The decoding formulation for the RNN-T model using phone as prediction is derived as below:
\begin{align}\label{eq-psd1}
    \textbf{w}^* &= \arg \max_\textbf{w}\{P(\textbf{w})p(\textbf{x}|\textbf{w})\} = \arg \max_\textbf{w}\{P(\textbf{w})p(\textbf{x}|\textbf{p}_\textbf{w}) \notag \\
        &= \arg \max_\textbf{w}\{P(\textbf{w})\frac{P(\textbf{x})p(\textbf{p}_\textbf{w}|\textbf{x})}{P(\textbf{p}_\textbf{w})}\}
        %&= argmax_w\{\frac{P(w)}{P(p_w)}max_{p_w}p(p_w|x)\}
\end{align}
where $\textbf{x}$ is the acoustic feature sequences, $\textbf{w}$ and $\textbf{p}_\textbf{w}$ are the word sequences and the corresponding phone sequences.  Equation \ref{eq-psd1} could be further simplified into below:
\begin{align}\label{eq-psd2}
    \textbf{w}^* &= \arg \max_\textbf{w}\{\frac{P(\textbf{w})}{P(\textbf{p}_\textbf{w})}\max_{\textbf{p}_\textbf{w}}p(\textbf{p}_\textbf{w}|\textbf{x})\}
\end{align}
We denote the standard decoding method as frame synchronous decoding (FSD) algorithm. When using the Viterbi beam search algorithm, FSD viterbi beam search could be transformed from the above equation~\ref{eq-psd2}  into:
%\vspace{-2mm}
\begin{align}\label{eq-psd3}
    \textbf{w}^* = \underset{\textbf{w}}{\arg \max}\{\frac{P(\textbf{w})}{P(\textbf{p}_\textbf{w})}\underset{\pi:\pi \in L', \beta(\pi_{1:T})=\textbf{p}_\textbf{w})}{\max}\{ \nonumber\\
    \prod_{t \not\in U}y^{t}_{\pi_{t}} \times \prod_{t \in U}y^{t}_{blank}  \}\}
\end{align}
%%\vspace{-1mm}
where $\pi$ is the possible alignment path. $L'$ is the CI phone set plus the blank symbol. $y_k^t$ represents the posterior probability of RNN-T output unit $k$ at time $t$. $U$ is a set including the time steps when $y^t_{blank}$ closes to one. The size of $U$ could be controlled by setting a threshold for blank labels' posterior $y^t_{blank}$. Since $\pi_{t}=blank$ won't change the corresponding phone sequences output $\beta{(\pi_{1:T})}$, assuming all competing alignment paths share the similar blank frames' positions, we could ignore the score of the blank frames. The below equation formulates our PSD algorithm on RNN-T lattice:
%\vspace{-2mm}
\begin{align}
    \textbf{w}^* = \underset{\textbf{w}}{\arg \max}\{\frac{P(\textbf{w})}{P(\textbf{p}_\textbf{w})}\underset{\pi:\pi \in L', \beta(\pi_{1:T})=\textbf{p}_\textbf{w})}{\max}
    \prod_{t \not\in U}y^{t}_{\pi_{t}}\}
\end{align}
In this way, the PSD method avoids redundant searches due to plenty of blank frames.  The PSD algorithm is summarized in Algorithm 1. We break the transducer lattice rule a little bit in decoding. One frame only outputs one phone label or blank \cite{tripathi2019monotonic}. 

To reduce the deletion errors caused by high blank label scores, we combine blank frames skipping strategy with blank label deweighting technology into Algorithm 1. We first deweight the blank scores by subtracting a deweighting factor in log domain. Then frames with blank scores more than predefined threshold would be filtered. The results in section 4.4 show the deweighting method could reduce the deletion errors significantly.   
%By skipping those frames, which are regarded as blank predictions, WFST decoding sequences' length reduces from $T$ to $T'$. 
By changing the threshold of blank frames, we could control how many blank frames would be skipped. 

\vspace{-3mm}
\begin{algorithm}[H]
\caption{PSD algorithm}
\label{alg:loop}
\begin{algorithmic}[1]
\Require{Features $\{\mathbf{x}_0,...,\mathbf{x}_{T-1}\}$, blank deweight value $\beta_{blank}$, blank threshold $\gamma_{blank}$}, Conv1d look-back M
\Ensure{ Predicted word sequences $\textbf{w}^*$}
    %\Statex
    %\State initialize phone sequences $y_{seq}$\gets [ ]
\State $\mathbf{y}_{in}={\rm Zeros(M)}, u=1, Q_{posteroir}=\{\}, \textbf{w}^{*}=\{\}$,
\State $\mathbf{h}_0^{pred}$=Predictor($\mathbf{y}_{in}$)
\For{ time $t \gets 0$ to $T-1$}
    %$posteriors\_table$ 
    \State $\mathbf{h}_t^{enc}$ = Encoder($\mathbf{x}_t$)
    ,$\mathbf{p}_{t,u}$ = Joint($\mathbf{h}_t^{enc}$, $\mathbf{h}_{u-1}^{pred}$)
    \State $\mathbf{p}_{t,u}(blank) = \mathbf{p}_{t,u}(blank) \times \beta_{blank}$
    \State $y^t_u=\arg \max \mathbf{p}_{t,u}$
    \If {$y^t_u \neq blank$}
        \State $u=u+1$
        \State $\mathbf{y}_{in} \gets [\mathbf{y}_{in}[1:],  y^t_u ]$
        \State $\mathbf{h}_u^{pred}$=Predictor($\mathbf{y}_{in}$)
    \EndIf
    \If {$\mathbf{p}_{t,u}(blank) \leq \gamma_{blank}$}
             
        \State Enqueue($Q_{posterior}$, $\mathbf{p}_{t,u}$)
        \State $\textbf{w}_{t,u}$=WFSTDecoding($LG$, $Q_{posterior}$)
        \State Enqueue($\textbf{w}^*$, $\textbf{w}_{t,u}$)
    \EndIf

\EndFor
\State \textbf{return} $\textbf{w}^*$
\end{algorithmic}
\end{algorithm}

\vspace{-5mm}
\subsection{Model Compression with SVD}
\label{ssec:subhead}
\vspace{-2mm}
We further reduce the model parameters using SVD. Since our parameters mainly come from the feed-forward projection layers in the DFSMN encoder, SVD is only used on these projection layers' weight matrices. Following the strategy in \cite{xue2013restructuring}, we first reduce the model size by SVD then fine-tune the compressed model to reduce the accuracy loss.

%\vspace{-5mm}

\section{Experiment}
\label{sec:page}
%\vspace{-4mm}
The experiments are conducted on an 18,000 hours in-car Mandarin speech dataset, which includes enquiries, navigations, and conversations speech collected from Tencent in-car speech assistant products. All the data are anonymized and hand transcribed. Development and Test set consist of 3382 and 6334 utterances, about 4 hours and 7 hours, respectively.

\vspace{-5mm}
\subsection{Model and Training Details}
\label{ssec:illust}
%%\vspace{-2mm}
Our model takes 40-dimensional power-normalized cepstral coefficients (PNCC) feature \cite{kim2016power} as input, which uses a 25ms window with a stride of 10ms. Adam optimizer, with an initial learning rate of 0.0005, is used to train the transducer model. Specaugment~\cite{park2019specaugment} with mask parameter (F = 20), and ten time masks with maximum time-mask ratio (pS = 0.05) is used as preprocessing. A 4-gram language model is trained using text data and additional text-only corpus. We have three different configurations for the large, medium, and small model's encoder. Predictors are all one layer CNN with different input dimensions according to the corresponding encoder size. The output units include 210 context-independent (CI) phones and the blank symbol. Transducer models are implemented with ESPnet \cite{watanabe2018espnet} toolkit. We first storage the predicted posterior probability matrices of CI phones. Then EESEN \cite{miao2015eesen} toolkit is used to process the posterior probabilities and gives the decoding results. Table 1 summarizes the model architecture details.
\vspace{-3mm}
\begin{table}[H]
  \caption{Details for Large, Medium and Small model}
   %\vspace{+3}
   \label{tab1}
  %\addtolength{\tabcolsep}{-3pt}
   \centering
    \begin{tabular*}{0.8\columnwidth}{llll}
  %\begin{tabular*}{\columnwidth}{llll}
    %\hline
    \toprule
    Model               & Large    & Medium & Small\\
    % \hline
    \midrule
                          
     \# Parameters           & 11M                & 4.5M          & 1.6M\\
                                  
   % \midrule
                 
      Encoder Dim                & 1024               & 512            & 400\\
      \# DFSMN layers         & 8                & 8             & 8 \\
      Joint Dim                  & 512                & 256             & 100 \\
     %  SVD                       & False              & False             & True \\
%\hline
    
    \bottomrule
  \end{tabular*}
  \vspace{-3mm}
\end{table}
%\vspace{-9mm}

\subsection{WER Results on Models}
\label{ssec:illust}
%\vspace{-2mm}
Table 2 shows the word error rate (WER) results of the conventional hybrid system, four RNN-T models with different sizes. They use the same language model. The hybrid system uses TDNN as the acoustic model with 2.5M parameters, which is comparable with our small transducer model.
By combining the end-to-end transducer model  with LG WFST decoder, we could surpass the hybrid system's performance and keep the flexibility of WFST to better customize the ASR system. Furthermore, Table~\ref{tab2} also shows results of the small model after SVD. With only 0.9M parameters, the SVD model with fine-tune could acheive 19.57\% CER, still better than hybird system. 
%\vspace{-5mm}
\vspace{-3mm}
\begin{table}[H]
  \caption{WER results on dev and test set}
  %\vspace{+4}
  \label{tab2}
  
  \centering
  %\hline
  \begin{tabular*}{0.85\columnwidth}{cccc}
    \toprule
    WER(\%)       &\# parameters        & Dev    & Test \\
    %\hline
    \midrule
                          
     Hybrid System   &2.5M        & 19.77              & 21.53          \\
                                  
    \midrule
                 
      Large Model   &11M            & 10.49               & 14.44            \\
      Medium Model  &4.5M            & 11.72               & 15.17              \\
      Small Model   &1.6M           & 14.12               & 18.11              \\
      + SVD fine-tune  & 0.9M           & 15.71                & 19.57 \\
     %  SVD                       & False              & False             & True \\
   %\hline
    
    \bottomrule
  \end{tabular*}
 \vspace{-3mm}
\end{table}

%\vspace{-7mm}
\subsection{RTF Results for PSD and FSD Algorithms}
\label{ssec:illust}
%\vspace{-3mm}
In this section, we would show the relationship between the blank rate and the corresponding threshold. Then we give the real-time factor (RTF), and WER results on our small model. We denote the blank rate $\alpha$ as follows:
\begin{align}
    \alpha = \frac{1}{T}size(U(\gamma_{blank}))
\end{align}
where $T$ is the sequence length and $U(\gamma_{blank})$ is the set including all blank frames:
\begin{align}
    U(\gamma_{blank}) = \{frames:\mathbf{p}_{t,u}(blank)>\gamma_{blank}\}
\end{align}
 The number of blank frames is controlled by the blank posterior probability threshold $\gamma_{blank}$. In decoding, all frames in the set $U$ would be skipped. When $\gamma_{blank}$ is larger than 1, no frames would be regarded as blank frames. In this case, the PSD algorithm would degrade to the FSD algorithm.     
\vspace{-3mm}
\begin{table}[H]
  \caption{Results with different threshold values}
  %\vspace{+3}
  \label{tab3}
  \setlength{\tabcolsep}{5pt}
  \centering
  %\hline
  \begin{tabular*}{0.97\columnwidth}{l|ll|ll|ll}
    \toprule
    \multicolumn{1}{c}{} & \multicolumn{2}{c}{} & \multicolumn{2}{c}{Speed} & \multicolumn{2}{c}{WER(\%)}\\
  %  Algorithm               & p  & $\alpha(\%)$ & Decoding  &Speed&  WER(\%) &\\\cmidrule{4-7}
    %               &   &  & Speed &  &  WER(\%)   & \\
    \midrule
   %\cmidrule{1-7}
     Method            & $\gamma_{blank}$  & $\alpha(\%)$ & RTF & S-RTF &  Dev   & Test\\
    %\hline
    \midrule
    FSD               & 1.0               & 0 &0.069 &0.053 &14.12 &18.11           \\
    \midrule
     PSD           & 0.95              & 77.08  &0.034 &0.017 &14.12 & 18.10                 \\
     PSD           & 0.85              & 80.73  &0.033 &0.017 &19.73 & 23.85               \\
     PSD           & 0.75              & 82.76  &0.032 &0.016 &36.79 & 41.28              \\
    % PSD           & adaptive          & 79.11  &0.033 &0.017 &14.08 & 18.05                 \\                             
    \bottomrule
  \end{tabular*}
 %\vspace{-3mm}
\end{table}
%\vspace{-3mm}
Table 3 gives a comparison of different threshold values and the corresponding results. Since PSD and FSD algorithms only have differences during WFST decoding time, we use RTF to represent the entire computation process, including transducer forward time and decoding time. S-RTF denotes the WFST search time. We conduct the experiments on a server with Intel(R) Xeon(R) E5 CPU for proof-of-concept. The results show that setting $\gamma_{blank}$ as $0.95$ would give a good balance between speed and accuracy.
%We could get the same accuracy while the decoding process yields an extra 3.1 times acceleration.

\subsection{Blank Frames Deweight}
\label{ssec:illust}
%\vspace{-2mm}
Following the strategy in \cite{7178778}, we don't normalize the phone label posteriors in decoding. We deweight the blank labels' posteriors to add a cost to deletion errors in decoding. Other label posteriors keep unchanged. We try to subtract a blank deweighting value on the log probability domain, which equals to divide a constant weight on blank labels' posterior probability. 
Figure 2 shows the deletion, substitution, and insertion errors for our small transducer model on the development set. We could always reduce the deletion errors by subtracting a higher deweighting number. However, too large deweighting values would increase the total WER. We tune the deweighting value on the development set and  using two as a deweighting value to get the best result.  

\begin{figure}[H]
  \centering
  \includegraphics[width=0.75\linewidth]{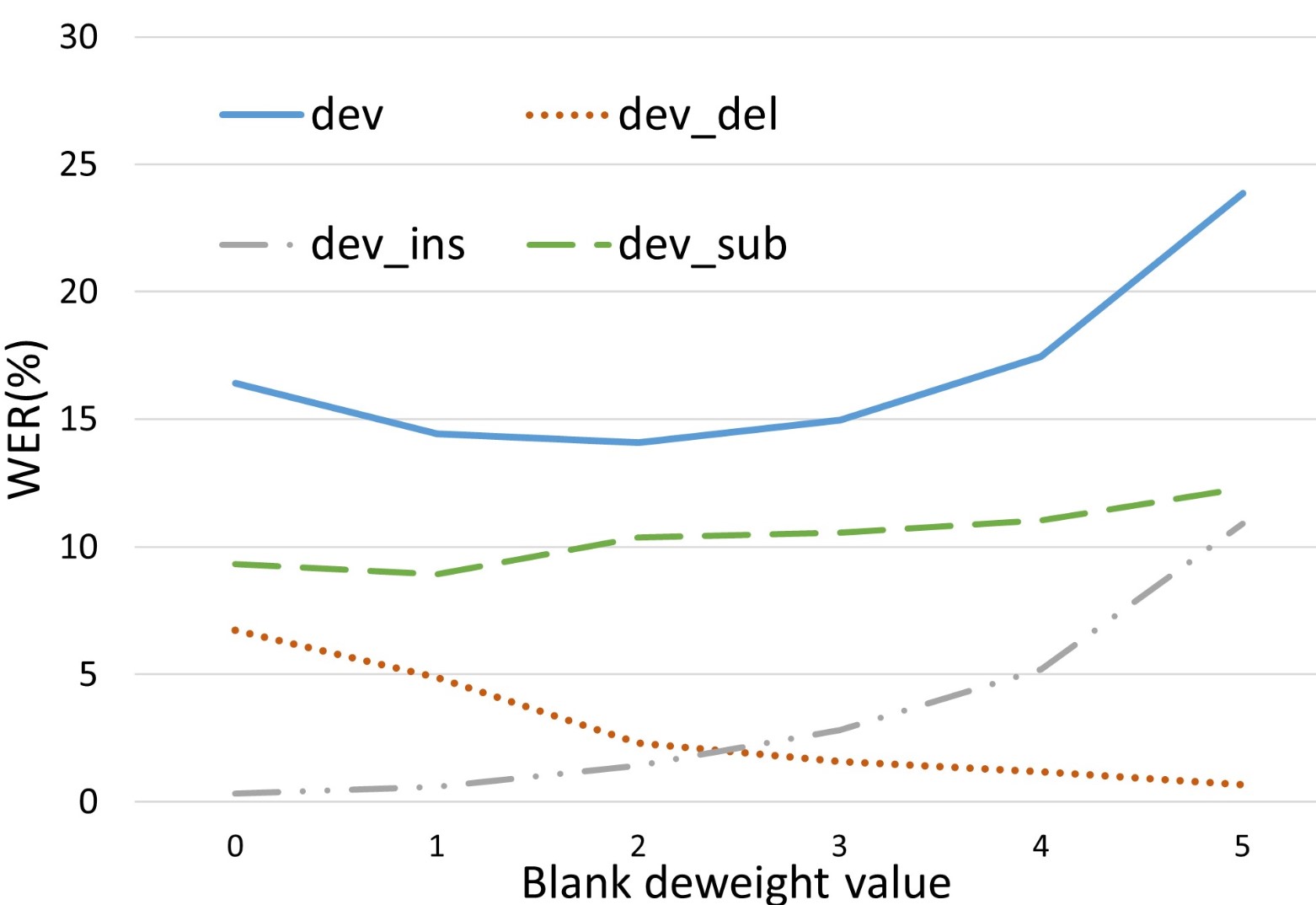}
  %%\vspace{-5mm}
  \caption{WER for different blank deweight value $\beta_{blank}$}
  \label{fig:speech_production2}
 % %\vspace{-5mm}
\end{figure}

\subsection{Performance on edge devices}

We also deploy our system on edge devices. Int8 quantization is used to reduce memory consumption and speed up inference. Note that, in order to trade off speech recognition accuracy and inference efficiency, only FSMN layers, which are parameter intensive, are quantized. Because our quantized model obtains similar accuracy as the results in Table~\ref{tab2}, we only report mean CPU usage and RTF of our small RNN-T model in Table~\ref{tab4}. From Table~\ref{tab4}, our proposed PSD method can significantly reduce CPU usage and RTF compared with FSD.      
\begin{table}[H]
\caption{On-device CPU usage and RTF results}
\label{tab4}
\centering
\scalebox{1.0}{
\begin{tabular}{llcc}
\toprule
Arm CPU                    &     & \begin{tabular}[c]{@{}c@{}}ARMv7\\ 4-core\end{tabular} & \begin{tabular}[c]{@{}c@{}}AArch64\\ 4-core\end{tabular} \\ \midrule
\multirow{2}{*}{CPU usage} & FSD & 48.5\%                                                            & 38.8\%                                                          \\
                           & PSD & 21.5\%                                                            & 6.2\%                                                           \\ \midrule
\multirow{2}{*}{RTF}       & FSD & 2.88                                                              & 2.66                                                            \\
                           & PSD & 0.55                                                              & 0.42                                                            \\ \bottomrule
\end{tabular}
}
\vspace{-3mm}
\end{table}
% \usepackage{multirow}

%%\vspace{-4mm}

%\vspace{-8mm}
\section{Conclusion}
\label{sec:copyright}
%\vspace{-3mm}
This paper introduces the pipeline of designing a highly compact speech recognition system for extremely low-resource edge devices. To fulfill the streaming attribute with low-computation and small model size constraints, we choose transducer lattice with  DFSMN encoder. LSTM predictor is replaced with a Conv1d layer to reduce the parameters and computation further. To keep the contextual and customizable recognition ability, we use CI phones as our modeling unit and bias the language model at the WFST decoding part. A novel PSD decoding algorithm based on transducer lattice is first proposed to speed up the decoding process. Also,  blank weight deweighting and SVD technologies are adopted to improve recognition performance. The proposed system shows a speech recognizer with few parameters which could realize streaming, fast, and accurate speech recognition. 

\ninept
\bibliographystyle{IEEEbib}
\bibliography{refs}

\end{document}